\documentclass[preprint]{vgtc}               

\graphicspath{{figures/}{pictures/}{images/}{./}} 

\usepackage{times}     
\usepackage{enumitem}

\usepackage{tabu}                      
\usepackage{booktabs}                  
\usepackage{lipsum}                    
\usepackage{mwe}                       

\usepackage{mathptmx}                  
\usepackage{colortbl}
\usepackage{adjustbox}
\usepackage{multirow}
\usepackage[table, svgnames]{xcolor}
\usepackage{tabularx}
\definecolor{NavyBlue}{RGB}{0,0,128}
\definecolor{darkgreen}{RGB}{131,197,190}

\newcommand{\heatmapColor}[1]{%
  \ifnum#1>95\cellcolor{darkgreen!100}\else%
  \ifnum#1>90\cellcolor{darkgreen!95}\else%
  \ifnum#1>85\cellcolor{darkgreen!90}\else%
  \ifnum#1>80\cellcolor{darkgreen!85}\else%
  \ifnum#1>75\cellcolor{darkgreen!80}\else%
  \ifnum#1>70\cellcolor{darkgreen!75}\else%
  \ifnum#1>65\cellcolor{darkgreen!70}\else%
  \ifnum#1>60\cellcolor{darkgreen!65}\else%
  \ifnum#1>55\cellcolor{darkgreen!60}\else%
  \ifnum#1>50\cellcolor{darkgreen!55}\else%
  \ifnum#1>45\cellcolor{darkgreen!50}\else%
  \ifnum#1>40\cellcolor{darkgreen!45}\else%
  \ifnum#1>35\cellcolor{darkgreen!40}\else%
  \ifnum#1>30\cellcolor{darkgreen!35}\else%
  \ifnum#1>25\cellcolor{darkgreen!30}\else%
  \ifnum#1>20\cellcolor{darkgreen!25}\else%
  \ifnum#1>15\cellcolor{darkgreen!20}\else%
  \ifnum#1>10\cellcolor{darkgreen!15}\else%
  \ifnum#1>5\cellcolor{darkgreen!10}\else%
  \ifnum#1>0\cellcolor{darkgreen!5}\else%
  \cellcolor{white}\fi\fi\fi\fi\fi\fi\fi\fi\fi\fi\fi\fi\fi\fi\fi\fi\fi\fi\fi\fi%
}

\usepackage[normalem]{ulem}

\onlineid{0}

\vgtccategory{Research}

\vgtcinsertpkg

\title{Exploring the Capability of LLMs in Performing Low-Level Visual Analytic Tasks on SVG Data Visualizations}

\author{Zhongzheng Xu\thanks{e-mail: zhongzheng\_xu@brown.edu}\\ %
        \scriptsize Brown University %
\and Emily Wall\thanks{e-mail: emily.wall@emory.edu}\\ %
     \scriptsize Emory University %
}

\abstract{
Data visualizations help extract insights from datasets, but reaching these insights requires decomposing high level goals into low-level analytic tasks that can be complex due to varying degrees of data literacy and visualization experience. Recent advancements in large language models (LLMs) have shown promise for lowering barriers for users to achieve tasks such as writing code and may likewise facilitate visualization insight. Scalable Vector Graphics (SVG), a text-based image format common in data visualizations, matches well with the text sequence processing of transformer-based LLMs. In this paper, we explore the capability of LLMs to perform 10 low-level visual analytic tasks defined by Amar, Eagan, and Stasko directly on SVG-based visualizations \cite{amar2005low}. Using zero-shot prompts, we instruct the models to provide responses or modify the SVG code based on given visualizations. Our findings demonstrate that LLMs can effectively modify existing SVG visualizations for some tasks like \textit{Cluster} but perform poorly on tasks requiring mathematical operations like \textit{Compute Derived Value}. We also discovered that LLM performance can vary based on factors such as the number of data points, the presence of value labels, and the chart type. Our findings contribute to gauging the general capabilities of LLMs and highlight the need for further exploration and development to fully harness their potential in supporting visual analytic tasks.
} 


\begin{document}


\firstsection{Introduction}

\maketitle
Data visualizations help people obtain meaningful insights from a dataset by encoding data into visual features such as length, position, shape, or color \cite{islam2019overview, dastani2002role}. Researchers have suggested that people often look at a chart with a high-level goal \cite{amar2005low}, such as: \textit{get an overview of the current stock market}. These high-level goals can be deconstructed into low-level tasks; following the example, it could include tasks such as \textit{finding the extreme values} of all-time stock prices or \textit{determining the range} of the stock prices over the course of a day \cite{amar2005low}. Amar, Eagan, and Stasko identified 10 such low-level visual analytic tasks that users commonly perform when exploring visualizations: \textit{Retrieving Values, Filter, Compute Derived Values, Find Extremum, Sort, Determine Ranges, Characterize Distribution, Find Anomalies, Cluster,} and \textit{Correlate} \cite{amar2005low}. For users to achieve their high-level goals, it may require them to decompose it into low-level tasks, the success of which may depend on several factors such as data literacy \cite{franconeri2022science} and experience levels \cite{burns2023we, grammel2010information, gupta2023belief}. 

\begin{figure}[ht]
  \begin{minipage}{\columnwidth}
    \centering
    \includegraphics[width=\linewidth]{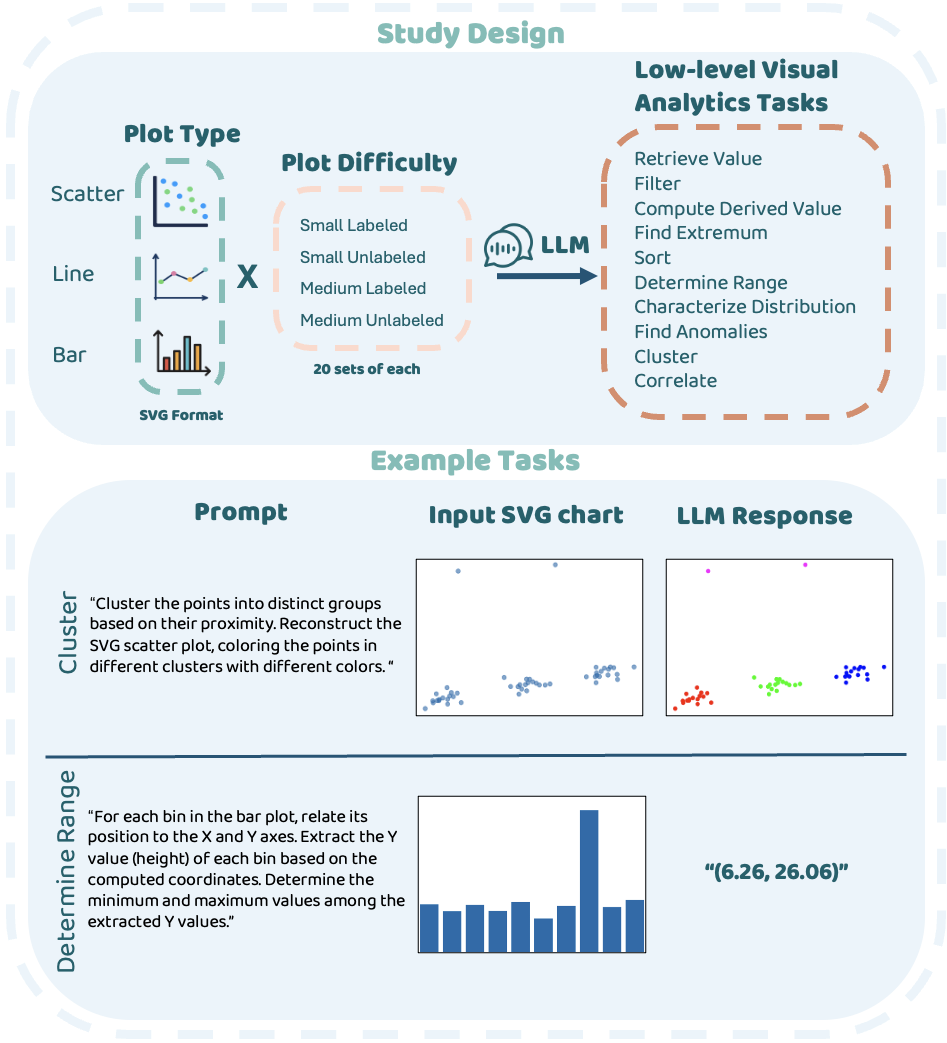}
    \caption{Illustration of the study design (top) and examples of the \textit{Cluster} and \textit{Determine Range} tasks (bottom).}
    \label{fig:design}
  \end{minipage}
\end{figure}

Recent advances in large language models (LLMs) have empowered people to complete difficult tasks such as reviewing legal documents \cite{deroy2023ready}, learning new languages \cite{jeon2023large}, and writing software code \cite{kazemitabaar2023novices} with greater ease. 
In the visualization community, researchers have incorporated LLMs into Natural Language Interfaces (NLIs) for chart creation \cite{tian2024chartgpt, maddigan2023chat2vis, hong2023conversational, han2023chartllama, dibia2023lida} and automatic chart summarization tools \cite{kantharaj2022chart, liew2022using}. 
A recent study found that LLMs are capable of understanding and modifying visualizations presented in Scalable Vector Graphics (SVG) format\cite{chen2023beyond}. 
Unlike raster-based images, SVG is a text-based image format that uses mathematical equations to define shapes, paths, and other elements through XML code \cite{quint2003scalable}. This text-based XML format of SVG images suggests that they may align well with transformer-based models' sequential text processing. Several studies have explored the capability of LLMs in understanding and manipulating SVG images \cite{cai2023leveraging, tseng2024keyframer}, but modifying existing visualizations is often a secondary focus, limited to editing styles such as changing background color or adding grid lines\cite{han2023chartllama, shi2023nl2color}.

In this paper, we explore an alternative usage of LLMs: to perform low-level visual analytic tasks directly on SVG-based visualizations. If LLMs can perform these low-level analytic tasks, it can reduce the barrier for end users to interact with data and gain insights. We thus conducted an exploratory evaluation of the ability of LLMs to understand and manipulate data visualizations using zero-shot prompts and instructed LLMs to perform tasks outlined by Amar et al. \cite{amar2005low} on a set of generated data visualizations. We observed that the LLM can effectively modify existing SVG visualizations for specific tasks involving pattern recognition such as \textit{Cluster} and \textit{Find Anomalies}. However, it performed poorly on tasks requiring complex mathematical operations, such as \textit{Compute Derived Values} and \textit{Correlation}. Our findings also highlight the LLM's capability to extract data point values directly from labeled SVG plots. We noted variations in performance based on the number of data points and the presence of value labels on the SVG charts. However, these effects were not consistent across all evaluated tasks. This baseline can provide valuable insights for future visualization researchers who are developing tools that incorporate LLMs 
and further contribute to gauging the general intelligence of LLMs. 


\section{Related Work}
Advancements in LLMs have sparked interest in bridging natural language processing and data visualization. Researchers have developed NLIs that enable the creation of data visualizations through natural language prompts \cite{tian2024chartgpt, maddigan2023chat2vis, hong2023conversational, han2023chartllama, dibia2023lida} and automatic chart summarization tools to generate captions for existing data visualizations \cite{kantharaj2022chart, liew2022using}. While some tools allow modifying existing visualizations, it is often limited to editing visual styles, such as colors or grid lines \cite{han2023chartllama}. Many of these tools require access to the original data, and few studies have explored using LLMs for direct chart manipulations. One example is the work by Shi et al. \cite{shi2023nl2color}, which focuses on using abstract natural language to refine color palettes in existing visualizations.

Evaluations of the capabilities of LLMs across various tasks have been conducted, including their math ability, coding skills, and machine translation, etc. \cite{chang2023survey, yuan2023large}. Several recent studies have explored the capability of LLMs in understanding and manipulating SVG images, such as changing element color in the image or editing animations \cite{cai2023leveraging, tseng2024keyframer}. In data visualization, Vázquez et al. evaluated the performance of several NLIs and libraries in creating charts using natural language \cite{vázquez2024llms}. Another work by Chen et al. used ChatGPT to complete an introductory data visualization course, revealing that LLMs can understand and manipulate SVG visualizations \cite{chen2023beyond}.

However, there is a lack of research systematically evaluating LLMs' capability to perform the low-level tasks that users typically carry out when exploring and interpreting data visualizations. 
We aim to shape our evaluation similar to that presented by Yuan et al.~\cite{yuan2023large} who evaluated LLM performance on elementary arithmetic tasks. That is, we will systematically explore LLMs' capability across low-level analytic tasks identified by Amar, Eagan, and Stasko \cite{amar2005low}. 

\section{Method}
In this study, we aim to evaluate the capability of LLMs to execute low-level visual analytic tasks on SVG-based data visualizations. We designed an experiment that assesses 3 chart types \{scatterplot, line chart, and bar chart\} \(\times\) 2 dataset sizes \{small, medium\} \(\times\) 2 labeling schemas \{labeled, unlabeled\} \(\times\) 10 visual analytic task prompts \{retrieve value, filter, compute derived value, find extremum, sort, determine range, characterize distribution, find anomalies, cluster, correlate\}.
We provided SVG and prompts as input to an LLM and subsequently evaluated task success. We detail the full methodology in this section and report on our findings in the next section.  

\smallskip

\textbf{Visualization Selection. }We evaluated the capability of LLMs to perform low-level visual analytic tasks on three chart types including scatterplots, line charts, and bar charts. These chart types are among the most widely used according to recent surveys \cite{li2020overview, quadri2021survey} and are also the commonly used in recent developments of NLIs for data visualization \cite{luo2021nvbench, maddigan2023chat2vis, han2023chartllama, tian2024chartgpt}. Extensive research has been conducted on the effectiveness of these chart types in supporting low-level analytic tasks, such as finding clusters \cite{quadri2022automatic, jeon2023clams}, identifying correlations \cite{miller2021accuracy, makowski2020methods}, and identifying trends \cite{albers2014task, wang2017line}. While these chart types do not cover the full spectrum of visualizations encountered in the wild, they are well-defined, goal-specific, and suitable for this particular study.

\smallskip

\textbf{Generation of Data and Stimuli. }We generated the SVGs using Python. The data generation pipeline was designed to create datasets suitable for performing certain low-level tasks, such as \textit{Cluster}. For scatterplots, points were initialized in multiple clusters separated by a random bias, with points within each cluster following a normal distribution and a specified x-y correlation factor. Line chart and bar chart datasets were generated by initializing points with a specified correlation factor (line chart) or random values within a range (bar chart) with random noise. Outlier points or bins were added to support the \textit{Find Anomalies} task for all three chart types.  The visualizations were plotted using the \textit{matplotlib} library.

Our preliminary experimentation revealed that LLM's performance could vary based on the number of data points and the presence of data encoding labels. Thus, we generated two dataset sizes \{small,medium\} and two labeling schemas \{labeled, unlabeled\} for each chart type. We created 20 unique datasets for each combination to account for dataset noise. For the task \textit{Characterize Distribution}, we generated an additional 20 datasets for each combination of chart type, dataset size, and labeling schema to evaluate the LLMs' ability to classify different distributions for scatterplots and bar charts, as the other datasets used a consistent distribution type. 
In total, we generated 400 unique stimuli (3 chart types $\times$ 2 dataset sizes $\times$ 2 labeling schemas $\times$ 20 random instantiations + 160 datasets for \textit{Characterize Distribution}). Within each chart type, the number of clusters, outliers, and initialization parameters were kept consistent. For labeled charts, we used the \textit{adjustText} library to ensure distinguishable labels \cite{flyamer2023phlya}. Full details on data generation and accompanying code are included in supplemental materials\footnote{\url{https://github.com/lebretou/SVG_taxonomy}}. 

\begin{figure}[ht]
  \begin{minipage}{\columnwidth}
    \centering
    \includegraphics[width=7cm]{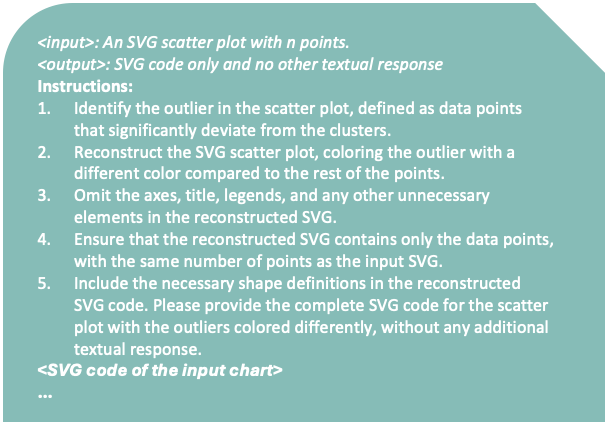}
    \caption{An example of the prompt for the task \textit{Find Anomalies}.}
    \label{fig:prompt}
  \end{minipage}
\end{figure}

\textbf{Prompt Engineering. }We employed OpenAI's \textit{gpt-4-turbo-preview}  model \cite{openai2023chatgpt4} to perform the 10 low-level  tasks on our generated datasets. It has been proven that prompts significantly impact the quality and correctness of the output of LLMs \cite{kervadec2023unnatural, gonen2022demystifying}; therefore, after preliminary experimentation, we designed our prompt to follow the format:
\textless Low-level task description\textgreater \textless SVG code\textgreater. 
In task descriptions, we instruct the model that it will answer questions or perform modification on a provided SVG code. Specifically, we followed the techniques and principles proposed in \cite{reynolds2021prompt, mishra2021reframing, chen2023unleashing} and explicitly defined the inputs and outputs of our tasks, as well as itemizing them into small steps. We then appended chart SVG code directly after the task descriptions. An example of the prompts used in our study is shown in Figure \ref{fig:prompt} for the \textit{Find Anomalies} task.

\smallskip
\textbf{Measuring Task Success. }To provide clarity on the specific tasks we evaluated, we present the tasks that the LLM performed below and the metric used to assess performance. 

Some low-level visual analytic tasks, such as \textit{Compute Derived Value}, are categories that encompass multiple specific tasks \cite{amar2005low}. For the purpose of our study, we selected one representative task for each category that aligns with the definition (e.g., for \textit{Compute Derived Value}, we chose to compute the mean only).
\begin{itemize}[noitemsep]
\item \textit{Retrieve Value}: Extract a list of coordinates (scatterplot, line chart) or bar heights (bar chart). 
\item \textit{Filter}: Create a plot with a subset of points (scatterplot, line chart) or bars (bar chart) that satisfy a given filtering criterion
\item \textit{Compute Derived Value}: Calculate the mean across the Y-axis (scatterplot, line chart, bar chart)
\item \textit{Find Extremum}: Highlight the points (scatterplot, line chart) or bars (bar chart) with minimum and maximum values in a different color
\item \textit{Sort}: Rearrange the bars (bar chart) in descending order based on their values
\item \textit{Determine Range}: Identify the range of values across the Y-axis (scatterplot, line chart, bar chart)
\item \textit{Characterize Distribution}: Identify the type of distribution exhibited by the points (scatterplot) or bins (bar chart) 
\item \textit{Find Anomalies}: Highlight anomalous points (scatterplot, line chart) or bars (bar chart) in a different color
\item \textit{Cluster}: Group points (scatterplot) into clusters, each represented by a different color
\item \textit{Correlate}: Add a fitted line that linearly fits the data points (scatterplot, line chart) in the plot
\end{itemize}

To measure the success of each task, we generated ground truth as answer keys for all the tasks. Although more complex evaluation metrics could be employed to measure partial success, we selected Exact Match (EM) \cite{chang2023survey} as our primary evaluation metric.

\begin{table}[!ht]
\centering
\setlength{\tabcolsep}{3pt}
\adjustbox{width=\columnwidth}{%
\begin{tabular}{llrrr}
  \hline
  \textbf{Task} & \textbf{Difficulty} & \multicolumn{1}{r}{\textbf{scatterplot}} & \multicolumn{1}{r}{\textbf{line chart}} & \multicolumn{1}{r}{\textbf{bar chart}} \\
  \hline
  \multirow{4}{*}{Retrieve Value} & small labeled & \heatmapColor{99} 99.1\% & \heatmapColor{100} 100\% & \heatmapColor{100} 100\% \\
   & small unlabeled & \heatmapColor{0} 0\% & \heatmapColor{0} 0\% & \heatmapColor{0} 0\% \\
   & medium labeled & \heatmapColor{86} 86.4\% & \heatmapColor{100} 100\% & \heatmapColor{100} 100\% \\
   & medium unlabeled & \heatmapColor{0} 0\% & \heatmapColor{0} 0\% & \heatmapColor{0} 0\% \\
  \hline
  \multirow{4}{*}{Filter} & small labeled & \heatmapColor{0} 0\% & \heatmapColor{55} 55\% & \heatmapColor{35} 35\% \\
   & small unlabeled & \heatmapColor{0} 0\% & \heatmapColor{40} 40\% & \heatmapColor{20} 20\% \\
   & medium labeled & \heatmapColor{0} 0\% & \heatmapColor{60} 60\% & \heatmapColor{5} 5\% \\
   & medium unlabeled & \heatmapColor{0} 0\% & \heatmapColor{65} 65\% & \heatmapColor{0} 0\% \\
  \hline
  \multirow{4}{*}{Compute Derived Value} & small labeled & \heatmapColor{0} 0\% & \heatmapColor{0} 0\% & \heatmapColor{0} 0\% \\
   & small unlabeled & \heatmapColor{0} 0\% & \heatmapColor{0} 0\% & \heatmapColor{0} 0\% \\
   & medium labeled & \heatmapColor{0} 0\% & \heatmapColor{0} 0\% & \heatmapColor{0} 0\% \\
   & medium unlabeled & \heatmapColor{0} 0\% & \heatmapColor{0} 0\% & \heatmapColor{0} 0\% \\
  \hline
  \multirow{4}{*}{Find Extremum} & small labeled & \heatmapColor{35} 35\% & \heatmapColor{70} 70\% & \heatmapColor{5} 5\% \\
   & small unlabeled & \heatmapColor{55} 55\% & \heatmapColor{80} 80\% & \heatmapColor{10} 10\% \\
   & medium labeled & \heatmapColor{35} 35\% & \heatmapColor{90} 90\% & \heatmapColor{30} 30\% \\
   & medium unlabeled & \heatmapColor{35} 35\% & \heatmapColor{85} 85\% & \heatmapColor{30} 30\% \\
  \hline
  \multirow{4}{*}{Sort} & small labeled & \cellcolor{white} N/A & \cellcolor{white} N/A & \heatmapColor{20} 20\% \\
   & small unlabeled & \cellcolor{white} N/A & \cellcolor{white} N/A & \heatmapColor{15} 15\% \\
   & medium labeled & \cellcolor{white} N/A & \cellcolor{white} N/A & \heatmapColor{5} 5\% \\
   & medium unlabeled & \cellcolor{white} N/A & \cellcolor{white} N/A & \heatmapColor{0} 0\% \\
  \hline
  \multirow{4}{*}{Determine Range} & small labeled & \heatmapColor{55} 55\% & \heatmapColor{95} 95\% & \heatmapColor{0} 0\% \\
   & small unlabeled & \heatmapColor{0} 0\% & \heatmapColor{0} 0\% & \heatmapColor{0} 0\% \\
   & medium labeled & \heatmapColor{5} 5\% & \heatmapColor{15} 15\% & \heatmapColor{0} 0\% \\
   & medium unlabeled & \heatmapColor{0} 0\% & \heatmapColor{0} 0\% & \heatmapColor{0} 0\% \\
  \hline
  \multirow{4}{*}{Characterize Distribution} & small labeled & \heatmapColor{0} 0\% & \heatmapColor{0} N/A & \heatmapColor{0} 0\% \\
   & small unlabeled & \heatmapColor{0} 0\% & \heatmapColor{0} N/A & \heatmapColor{0} 0\% \\
   & medium labeled & \heatmapColor{0} 0\% & \heatmapColor{0} N/A & \heatmapColor{0} 0\% \\
   & medium unlabeled & \heatmapColor{0} 0\% & \heatmapColor{0} N/A & \heatmapColor{0} 0\% \\
  \hline
  \multirow{4}{*}{Find Anomalies} & small labeled & \heatmapColor{80} 80\% & \heatmapColor{100} 100\% & \heatmapColor{85} 85\% \\
   & small unlabeled & \heatmapColor{60} 60\% & \heatmapColor{95} 95\% & \heatmapColor{95} 95\% \\
   & medium labeled & \heatmapColor{65} 65\% & \heatmapColor{95} 95\% & \heatmapColor{70} 70\% \\
   & medium unlabeled & \heatmapColor{45} 45\% & \heatmapColor{100} 100\% & \heatmapColor{75} 75\% \\
  \hline
  \multirow{4}{*}{Cluster} & small labeled & \heatmapColor{95} 95\% & \cellcolor{white} N/A & \cellcolor{white} N/A \\
   & small unlabeled & \heatmapColor{95} 95\% & \cellcolor{white} N/A & \cellcolor{white} N/A \\
   & medium labeled & \heatmapColor{95} 95\% & \cellcolor{white} N/A & \cellcolor{white} N/A \\
   & medium unlabeled & \heatmapColor{95} 95\% & \cellcolor{white} N/A & \cellcolor{white} N/A \\
  \hline
  \multirow{4}{*}{Correlate} & small labeled & \heatmapColor{0} 0\% & \heatmapColor{0} 0\% & \cellcolor{white} N/A \\
   & small unlabeled & \heatmapColor{0} 0\% & \heatmapColor{0} 0\% & \cellcolor{white} N/A \\
   & medium labeled & \heatmapColor{0} 0\% & \heatmapColor{0} 0\% & \cellcolor{white} N/A \\
   & medium unlabeled & \heatmapColor{0} 0\% & \heatmapColor{0} 0\% & \cellcolor{white} N/A \\
  \hline
\end{tabular}%
}
\caption{The performance of the LLM on low-level visual analytic tasks across different chart types and difficulty levels. The percentage scores indicate the Exact Match scores between the LLM's output and the ground truth answer key. "N/A" indicates that the task was not tested for a particular chart type due to its inapplicability.}
\label{tab:results}
\end{table}

\section{Result}
In this section, we present the results of the LLM's performance on each low-level visual analytic task using the Exact Match metric. Table \ref{tab:results} provides an overview of the scores across different chart types and dataset size and labeling schemas. We structure this section by discussing the results for each low-level task individually, comparing the performance across scatterplots, line charts, and bar charts. Note that many task success rates will fall in increments of $5\%$ given that we measure exact match across 20 random dataset instantiations, with the exception of the \textit{retrieve value} task, for which we compute the exact match for each of the N respective data points.

\begin{figure}[ht]
  \begin{minipage}{\columnwidth}
    \centering
    \includegraphics[width=7cm]{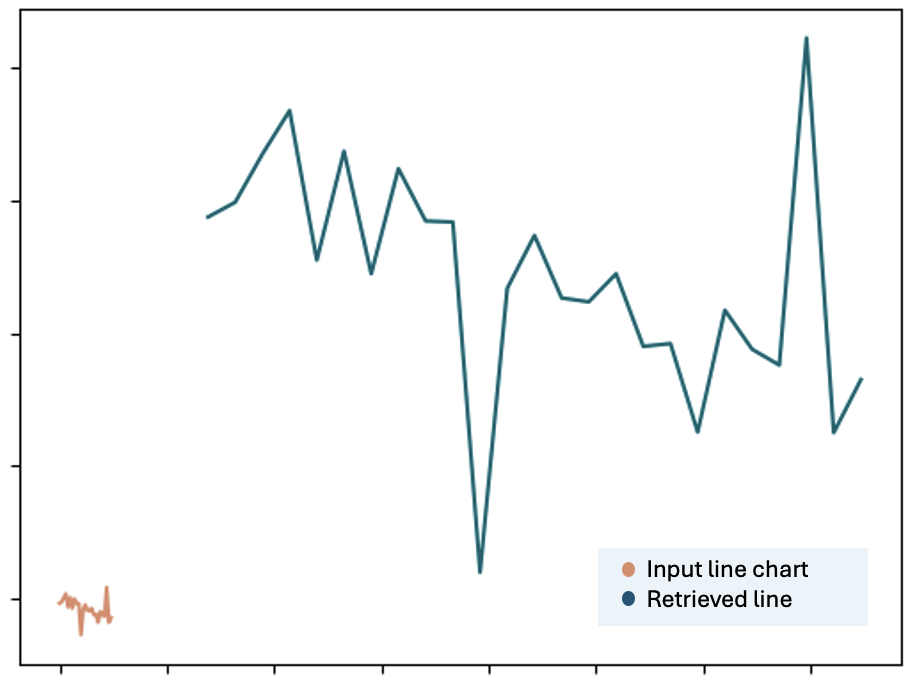}
    \caption{Comparison of retrieved data points (teal) extracted by the LLM directly from SVG position definitions and the original data points (orange).}
    \label{fig:retrieval}
  \end{minipage}
\end{figure}

\smallskip
\textit{1. Retrieve Value}: The LLM achieved 100\% accuracy for line charts and bar charts with labeled data values. For scatterplots, the accuracy was slightly lower due to missed and incorrectly identified data points in some responses.
When charts lacked data value labels, the LLM couldn't retrieve exact values for all three chart types, with responses often containing imaginary data points. Scatterplots were the most affected.
However, for many line chart and bar chart responses, the retrieved points, when plotted, resulted in shapes similar to the original data points. Pearson correlation revealed that the LLM retrieved points resulting in the same shape as the original chart in 17/20 "small" and 10/20 "medium" cases for line charts, and 15/20 "small" and 7/20 "medium" cases for bar charts, achieving a 99\% correlation factor. The remaining responses were hallucinations with no correlation to the original data.
Inspection revealed that the LLM extracted x,y coordinates from the SVG code's point position definitions, despite the prompt stating not to do so (as SVG pixel coordinates don't equate to point positions on the defined scales). Figure \ref{fig:retrieval} compares the retrieved data points, extracted directly from SVG position definitions, against the original data points.

\smallskip
\textit{2. Filter}: The LLM had the highest accuracy for line charts, followed by bar charts, with no successful cases for scatterplots. For line charts, accuracy was higher with more data points ("medium" cases) and unlabeled data points.

Scatterplots suffered from hallucination, with the LLM returning a subset of points that didn't meet the specified filtering criteria or sometimes including all original points without filtering. Some incorrect responses for line charts and bar charts returned a subset of original points that didn't meet the desired filtering conditions.

\smallskip
\textit{3. Compute Derived Value}: The LLM had no success, regardless of whether the points were labeled with their exact values or not. The returned values were generally hallucinations. We discuss this finding in greater detail in Section~\ref{sec:discussion}.

\smallskip
\textit{4. Find Extremum}: The LLM's accuracy was highest for line charts, followed by scatterplots and bar charts. Interestingly, in some cases, the LLM performed better when data values were not labeled. For line and bar charts, accuracy improved with more data points. In several incorrect responses across all chart types, the LLM highlighted only one extremum or sometimes local extrema instead of global extrema.

\smallskip
\textit{5. Sort}: The LLM achieved the highest accuracy for bar charts in the "small labeled" cases, followed by "small unlabeled" and "medium labeled" cases, but failed to correctly sort bars for "medium unlabeled" cases. Inspection of failed responses revealed that the LLM attempted to rearrange the bars, but not always in the specified "descending order," sometimes outputting bars in ascending order instead.

\smallskip
\textit{6. Determine Range}: The LLM had the best accuracy for line charts, followed by scatterplots, and it was not able to determine the range for bar charts. When the data points were not labeled, the LLM failed to determine the range for all three chart types.
As with the \textit{Retrieve Value} task, the LLM gave imaginary points and mistakenly extracted (x,y) point positions defined in the SVG code. 

\smallskip
\textit{7. Characterize Distribution}: The LLM had no success for scatterplots or bar charts. The data were generated according to normal, binomial, or uniform distributions. The LLM responded with hallucinations like "positive correlation" for all 20 plots. However, when the correct distribution options were explicitly included in the prompt (e.g., "The answer should be one of [normal, uniform, binomial]."), the LLM's accuracy improved to 37.5\% for scatterplots and 35\% for bar charts, reflecting chance accuracy. 

\smallskip
\textit{8. Find Anomalies}: The LLM had the highest accuracy for line charts, followed by bar charts and scatterplots. Notably, for line charts and bar charts, the performance was comparable when data values were labeled and not labeled, though worse when not labeled for scatterplots. 
In the failed responses, the LLM sometimes highlighted incorrect points, such as a point among a cluster, and sometimes it would only highlight one outlier instead of two for the ``medium" cases.

\smallskip
\textit{9. Cluster}: The LLM successfully highlighted different clusters in colors with an accuracy of 95\% for all cases. 
In the failed responses, the LLM provided incorrect clustering results with more clusters than the number specified in the prompt.

\smallskip
\textit{10. Correlate}: The LLM failed to add a properly fitted line to the chart for both scatterplots and line charts. Instead, most of the responses included an arbitrarily placed line, indicating hallucination.
Notably, this was the only task for which the LLM refused to respond with an SVG code output, stating, "I am unable to modify SVG files."

\section{Discussion and Conclusion}
\label{sec:discussion}
In this section, we discuss what we expected and did not expect from the performance of the LLM in performing low-level visual analytic tasks, including analysis of factors that influenced the LLM's performance. We synthesize the implications of our findings for the field of data visualization, then identify limitations of our approach and future research directions.

\smallskip
\noindent\textbf{The Expected.} The LLM performed poorly when complex mathematical operations were required. For example, retrieving unlabeled data point values requires subtraction and division based on axis labels. Similarly, computing the mean across an axis involves addition, counting, and division. This aligns with previous research on LLMs' limited capability in handling long expressions, especially when directly outputting answers \cite{yuan2023large}. However, there is promise in integrating scripting approaches alongside LLMs, as ChatGPT does with its web interface (though not in the API), discussed further in Implications. 

\smallskip
\noindent\textbf{The Unexpected.} Contrary to expectations, fewer data points did not always lead to higher accuracy. For tasks like \textit{Finding Extremum}, the LLM had higher accuracy for line and bar charts with more data points. Likewise, data point labels do not always affect performance; in \textit{Finding Extremum}, the LLM performed comparably regardless of labeling, indicating its ability to extract information from SVG shape definitions alone. Surprisingly, despite accurately retrieving values from labeled charts, the LLM's performance in determining value ranges was poor. However, the LLM's high accuracy in clustering scatterplot points suggests its adeptness in recognizing and grouping patterns.
\smallskip

\noindent\textbf{Implications.} The differences in LLMs' capability to perform low-level visual analytic tasks offer valuable insights. Unlike previous tools like NLIs that require data to create plots, LLMs can directly perform low-level tasks on existing visualizations without the original data. 
However, relying solely on LLMs' inherent abilities may not suffice, as they can ``lose track" of themselves in intermediate steps \cite{mishra2021reframing, chen2023unleashing}. Future research could explore bolstering training of sub-tasks where LLMs fail to perform accurately, as fine-tuning LLMs has been shown to substantially enhance their abilities in trained domains, suggesting the possibility of developing reliable tools that incorporate LLMs for tasks like clustering points in a scatterplot \cite{ding2023parameter, chen2023beyond}.

Combining LLMs with other techniques could likewise enhance their effectiveness and accuracy in low-level visual analytic tasks. ChatGPT's web interface, which supports running code within the session, demonstrates how integrating LLMs with other tools can improve accuracy in tasks like \textit{Determine Range} by executing Python code on coordinates extracted from data labels \cite{openai2023chatgpt4}. Finally, our motivation began with the hope that LLMs could facilitate end-users accomplishing high-level goals of data comprehension and insight generation from data visualizations. For these findings to be put into practice, we must understand how to compose the low-level tasks that have high accuracy into high-level visual analytic goals, drawing inspiration from the decomposition approach attempted by Tian et al. in their NLI for chart creation \cite{tian2024chartgpt}.


\smallskip
\noindent\textbf{Limitations and Future Work.}

\noindent Our work is limited by the small and homogenous set of charts evaluated. Future studies could test a wider array of chart types and variations. Additionally, our evaluation used only OpenAI's \textit{gpt-4-turbo-preview} \cite{openai2023chatgpt4}. Future research could compare the performance across different models. Moreover, we utilized zero-shot prompts, depending solely on the model's built-in knowledge. As \cite{chen2023beyond} indicates, in-context learning and fine-tuning might enhance LLM performance on specialized tasks.

\smallskip
\noindent\textbf{Conclusion. }We conducted an exploratory evaluation of an LLM's potential in performing low-level visual analytic tasks on SVG data visualizations. LLMs are capable of executing tasks like identifying outliers and clustering points in scatterplots, but struggle with tasks requiring complex mathematical operations. Performance varies based on the number of data points, value label presence, and chart type.
Our findings contribute to the growing research on LLMs in data visualization and highlight the need for further exploration and development to harness their potential in supporting visual analytic tasks.


\acknowledgments{
This work was partially supported by NSF award IIS-2311574.}

\bibliographystyle{abbrv-doi}

\bibliography{template}
\end{document}